\documentclass[11pt]{article}
 
\makeatletter
\@addtoreset{equation}{section}
\makeatother

\def\text#1{\mbox{\rm #1}}
\begin{document}
\author{{\bf Delia Ionescu}\\
{\it
Department of Mathematics, Technical University of Civil Engineering, Bucharest}\\ 
{\it Bd. Lacul Tei, No. 124, Bucharest, Romania; E-mail: dionescu@hidro.utcb.ro}}
\title{{\LARGE {\bf The  Electrogravitational Field of an Electrically Charged Mass Point 
and the Causality Principle in RTG }}}
\date{}
\maketitle
\begin{abstract}
\parbox{13cm}
{In this paper, I determine the electrogravitational field produced by a charged mass point 
according to the Relativistic Theory of Gravitation. The Causality Principle in the Relativistic 
Theory of Gravitation will play
a very important part in finding this field. The analytical form and the domain of definition,
i.e the gravitational radius of the obtained solution, differ from that given by Einstein's General 
Relativity Theory.
\smallskip

{\bf Keywords}: Relativistic Theory of Gravitation, Electrogravitational Fields, 
Causality Principle.}
\end{abstract}

\smallskip

\medskip

\bigskip %\hsize 160mm\vsize 230mm

\section{Introduction}

\smallskip

The purpose of this paper is to find the electrogravitational field produced
by a charged mass point in the framework of the Relativistic Theory of
Gravitation (RTG).

In Section 2, I present the basics of this new relativistic theory of
gravitation elaborated by Logunov and his co-workers (see Logonov \& Mestvirishvili 
1989 and Logunov 1997). In RTG, a
gravitational field is determined unambiguously solving the complete system of
RTG's Eqs. The Causality Principle (CP) in RTG, permits the selection of
those solutions of RTG's Eqs. which can have physical meaning.

In Section 3, I present a formulation of the basic laws of electromagnetism
in the vacuum, in the presence of the gravitational field. The presentation
is based on the fact that the behaviour of an electromagnetic field can be
formulated in a four dimensional manifold without supplementary mathematical
structure (see Truesdell \& Toupin 1960, Sections 266-275 and
also So\'{o}s 1992 Sections 1-3).

The problem of finding the field of an electrically charged mass point was
solved in the framework of Einstein's General Relativity Theory (GRT) by
Nordstr\"{o}m and Jeffrey (see Wang 1979, Section 56). 
At the beginning of Section 4, I
present the solution obtained by them. But as was shown by Logunov and his
co-workers, in GRT, one can not obtain an unique solution of the problem
without introducing prior assumptions (see Logonov \& Mestvirishvili  
1989, Section12). Besides, the
gravitational radius of the source point, as function depending on $q^{2}$
and $m^{2}$, where $q$ and $m$ representing the electric charge and the mass
of the point source, has a discontinuity in $q^{2}=m^{2}.$

In RTG, the considered problem was firstly analysed by Karabut \& Chugreev (1987),
but only assuming $m^{2}\geq q^{2}$. In Section 5, I present this
solution. I verify also that this solution satisfies CP in RTG, so, it's a
solution which has physical sense.

So\'{o}s and me (2000) have reanalyzed the problem in RTG, considering
also the possibility $q^{2}>m^{2}.$ It's important to analyse this case
because for the electron just this variant is true. The analytical form of
the solution found by us, as well as its domain of definition, i.e. the
gravitational radius $r_{g}$, depend essentially on the relation existing
between $q^{2}$ and $m^{2}$. But, I'll show in Section 6 that this solution
doesn't fulfill CP in RTG. So, this solution can't be an acceptable solution
in this theory.

I determine in Section 6, the unique solution of electrogravitational field
produced by a charged mass point according to RTG. The obtained solution has
the same analytical form for all order relations between $q^{2}$ and $m^{2}$%
. The gravitational radius depend on this relation but it's a continuous
function depending on $q^{2}$ and $m^{2}.$

\section{RTG's equations and the Causality Principle in RTG}

\smallskip

RTG was constructed by Logunov and his co-workers (see Logonov \& Mestvirishvili 
1989 and Logunov 1997) as a field
theory of the gravitational field within the framework of Special Relativity
Theory (SRT). The Minkowski space-time is a fundamental space that
incorporates all physical fields, including gravitation. The line element of
this space is:

\hspace{5cm} 
\begin{equation}
d\sigma ^{2}=\gamma _{mn}(x)dx^{m}dx^{n},
\end{equation}
where $x^{m},$ $m=$1, 2, 3, 4, is an admissible coordinate system in the
underlying Minkowski space-time; $\gamma _{mn}(x)$ are the components of the
Minkowskian metric in the assumed coordinate system.

The gravitational field is described by a second order symmetric tensor $%
\phi ^{mn}(x)$, owing to the action of which un effective Riemannian
space-time arises.

One of the basic assumption of RTG\ tells us that the behaviour of matter in
the Minkowskian's space -time with metric $\gamma _{mn}(x)$, under the
influence of the gravitational field $\phi ^{mn}(x),$ is identical to its
behaviour in the effective Riemannian space-time with metric $g_{mn}(x)$,
determined according to the rules:

\begin{equation}
\tilde{g}^{mn}=\sqrt{-g}\mbox{\rm  }g^{mn}=\sqrt{-\gamma }\mbox{\rm  }\gamma
^{mn}+\sqrt{-\gamma }\mbox{\rm  }\phi ^{mn}\mbox{\rm  },\mbox{\rm  }g=\det
(g_{mn})\mbox{\rm  },\mbox{\rm  }\gamma =\det (\gamma _{mn}).
\end{equation}

Such interaction of the gravitational field with matter was termed the
geometrization principle of RTG.

The behaviour of the gravitational field is governed by the following
differential laws of RTG:

\begin{eqnarray}
R_{n}^{m}-\frac{1}{2}\delta _{n}^{m}R+\frac {m_g^2}{2}
\left(\delta^m_n+g^{mk}\gamma_{kn}-\frac {1}{2} \delta^m_ng^{kl}\gamma_{kl}
\right)&=&8\pi T_{n}^{m}\mbox{\rm  },%
 \\
D_{m}\tilde{g}^{mn} &=&0\mbox{\rm  , \quad }m,n,p=1,2,3,4.
\end{eqnarray}

Here $R_{n}^{m}$ is Ricci's tensor corresponding to $g_{mn}$ ,
$R=R_{m}^{m}$ is the scalar curvature,
$\delta
_{n}^{m}$ are Kronecker's symbols and $T_{n}^{m}$ denotes the
energy-momentum tensor of the sources of the gravitational field. In (2.4) $%
D_{m}$ is the operator of covariant differentiation with respect to the
metric $\gamma _{mn}$. Eqs. (2.3), (2.4) are covariant under arbitrary
coordinate transformations with a nonzero Jacobian.
All the field variables in the RTG depend on universal space-time
coordinates of the Minkowski space. The presence of mass terms
in Eq. (2.3) allows unambiguously determining the space-time
geometry and the density of the gravitational field energy-momentum
tensor in the absence of matter.
 Eqs. (2.4) tell us that a gravitational field
can have only the spin states 0 and 2. In Logonov \& Mestvirishvili
(1989), this represents one of the basic assumption of RTG. In Logunov (1997), these Eqs. which
determine the polarization states of the field, are consequences of the fact
that the source of the gravitational field is the universal conserved
density of the energy-momentum tensor of the entire matter including the
gravitational field.
The graviton mass essentially affects the Universe's evolution and changes the character of the gravitational
collapse.

In the present paper, because the graviton mass is negligibly small, we omitt the mass terms in Eq. (2.3).
Hereafter, we use the relativistic system of units.

Eqs. (2.4) can be written in the following form (see Logonov \& Mestvirishvili 1989, Appendix 1):

\qquad \qquad \qquad \qquad \qquad \qquad \qquad
$D_{m}\tilde{g}^{mn}=\tilde{g}^{mn},_{m}+\gamma _{mp}^{n}\tilde{g}^{mp}=0$,
\qquad \qquad \qquad \qquad\qquad(2.4') where $\gamma _{mp}^{n}$ are the components of the metric connection
generated by $\gamma _{mn}$ and the comma is the derivation relative to the
involved coordinate.The causality principle (CP) in RTG is presented and
analysed by Logunov (1997), Section 6.

According to CP any motion of a pointlike test body must have place within
the causality light cone of Minkowski's space-time. According to Logunov's
analysis CP will be satisfied if and only if for any isotropic Minkowskian
vector $u^{m}$ , i.e. for any vector $u^{m}$ satisfying the condition:

\begin{equation}
\gamma _{mn}u^{m}u^{n}=0,
\end{equation}
the metric of the effective Riemannian space-time satisfies the restriction:

\begin{equation}
g_{mn}u^{m}u^{n}\leq 0
\end{equation}

According to CP of RTG\ only those solutions of the system (2.3), (2.4) can
have physical meaning which satisfies the above restriction.

It's important to stress the fact that CP\ in the above form can be
formulated only in RTG, because only in this theory, the space-time is
Minkowskian and the gravitational field is described by a second order
symmetric tensor field $\phi _{mn}(x)$, $x^{m}$ being the admissible
coordinates in the underlying Minkowskian space-time, $x^{1},x^{2},x^{3}$
being the space-like variables and $x^{4}$ being the time-like variable.

\smallskip

\section{Electromagnetic field equations in RTG}

\smallskip

The theory of electromagnetism is a very elegant and formally simple part of
physics. The two principles of conservation set up as the basis for this
theory, are the conservation of charge and the conservation of magnetic flux
(see Truesdell \& Toupin 1960, Sections 266-270, and So\'{o}s 1992, Sections 1-3).
 These two laws of
conservation can be formulated in a four dimensional manifold independent of
any geometry of space-time. The differential form of the so-called
Maxwell-Bateman laws is (see So\'{o}s 1992, Sections 1-3 ):

\begin{eqnarray}
\Psi _{,m}^{mn} &=&K^{n} \\
F_{,m}^{mn} &=&0\mbox{\rm  , \quad }m,n,p=1,2,3,4,
\end{eqnarray}
where $F^{mn}$ is a contravariant axial 2-vector density representing the
electromagnetic field, $\Psi ^{mn}$ is a contravariant 2-vector density
representing the electromagnetic induction and $K^{n}$ is a contravariant
1-vector density representing the electric current and the charge density.
All the involved fields in Eqs. (3.1), (3.2), depend on an admissible
coordinate system $(x^{n})$ in the four dimensional manifold.

The conception of conservation as formulated here have a topological
significance, being independent of the mathematics of length, time and
angles. To interpret Eqs. (3.1), (3.2) in the familiar language of
electromagnetic theory, a metric structure must be considered on the four
dimensional manifold.

We assume now that our manifold is the Minkowski space-time of RTG. In this
case we suppose that ($\alpha ,\beta ,\delta =1,2,3$): 
\begin{equation}
F^{\alpha \beta }=\varepsilon ^{\alpha \beta \delta }E_{\delta },\mbox{\rm  }%
F^{\alpha 4}=B^{\alpha },\mbox{\rm  }\Psi ^{\alpha \beta }=\varepsilon
^{\alpha \beta \delta }H_{\delta },\mbox{\rm  }\Psi ^{\alpha 4}=-D^{\alpha },%
\mbox{\rm  }K^{\alpha }=j^{\alpha },\mbox{\rm  }K^{4}=\rho .\mbox{\rm  }
\end{equation}
In the above Eqs. $\varepsilon ^{\alpha \beta \delta }$ are the
contravariant Ricci's permutation symbols, $E_{\delta }$ are the covariant
components of the electric field and $B^{\alpha }$ the contravariant
components of the magnetic field, corresponding to the selected Minkowskian
coordinate system which may be inertial or not. Also, $H_{\delta }$ are the
covariant components of the magnetic induction and $D^{\alpha }$ are the
contravariant components of the electric induction, whereas $j^{\alpha }$
and $\rho $ represent the contravariant components of the electric current
and the density of the electric charge, respectively.

As a fundamental hypothesis of the relativistic electromagnetism in vacuum
and excluding gravitation we assume that $F^{mn}$ and $\Psi ^{mn}$ are
connected by the Maxwell-Lorentz aether relation (see  Truesdell \& Toupin 1960,
Section 280, and So\'{o}s 1992, Section 24):

\begin{equation}
\Psi ^{mn}=\sqrt{-\gamma }\gamma ^{mp}\gamma ^{nq}\hat{F}_{pq}\mbox{\rm  },%
\mbox{\rm 
}\hat{F}_{pq}=\frac{1}{2}\varepsilon _{mnpq}F^{mn}\mbox{\rm  },\mbox{\rm  }%
m,n,p,q=1,2,3,4,
\end{equation}
where $\varepsilon _{mnpq}$ are Ricci's covariant permutation symbols and $%
\hat{F}_{pq}$ is the dual of $F^{mn}$, being a skew-symmetric absolut
tensor. From (3.4), in a Minkowskian inertial frame we get the familiar
relations characterizing the vacuum:

\begin{equation}
D^{\alpha }=E_{\alpha }\mbox{\rm  , }H_{\alpha }=B^{\alpha }.
\end{equation}

Let us assume now the presence of the gravitational field. Eqs. (3.1), (3.2)
rest also true. Taking into account the geometrization principle of RTG , I
assume that in the presence of the gravitational field the Maxwell-Lorentz
aether relation takes the following form:

\begin{equation}
\Psi ^{mn}=\sqrt{-g}g^{mp}g^{nq}\hat{F}_{pq}\mbox{\rm  },\mbox{\rm  }\hat{F}%
_{pq}=\frac{1}{2}\varepsilon _{mnpq}F^{mn}\mbox{\rm  },\mbox{\rm  }%
m,n,p,q=1,2,3,4,
\end{equation}

Introducing (3.6) into (3.1), (3.2) and taking into account that for any
skew-symmetric tensor $X^{mn}$ we have the formula $\bigtriangledown
_{m}X^{mn}=\frac{1}{\sqrt{-g}}\frac{\partial }{\partial x^{m}}\left( \sqrt{-g%
}X^{mn}\right) ,$ $\bigtriangledown _{m}$being the operator of covariant
differentiation with respect to the metric $g_{mn}$ , the laws describing
the behaviour of an electromagnetic field in RTG can be written in the form:

\begin{equation}
\bigtriangledown _{m}\hat{F}^{mn}=J^{n},
\end{equation}

\begin{equation}
\bigtriangledown _{i}\hat{F}^{mn}+\bigtriangledown _{m}\hat{F}%
^{ni}+\bigtriangledown _{n}\hat{F}^{im}=0,
\end{equation}
where $J^{n}\equiv \frac{K^{n}}{\sqrt{-g}}.$

To get the energy-momentum tensor $T_{n}^{m}$ of the electromagnetic field
(in vacuum) in RTG we start with its expression in the relativistic
electromagnetism excluding gravitation (see Moller 1972, Section 7):

\begin{equation}
\sqrt{-\gamma }T_{n}^{m}=-\frac{1}{4\pi }\hat{F}_{np}\Psi ^{mp}+\frac{1}{%
16\pi }\mbox{\rm  }\hat{F}_{pq}\Psi ^{pq}\delta _{n}^{m}.\mbox{\rm  }
\end{equation}

Using again the geometrization principle of RTG and the assumed
Maxwell-Lorentz aether relations (3.6), we can conclude that in RTG, $%
T_{n}^{m}$ are given by:

\begin{equation}
T_{n}^{m}=-\frac{1}{4\pi }\hat{F}_{np}\hat{F}^{mp}+\frac{1}{16\pi }%
\mbox{\rm
}\hat{F}_{pq}\hat{F}^{pq}\delta _{n}^{m}\mbox{\rm  , }\hat{F}%
^{mp}=g^{mi}g^{pj}\hat{F}_{ij}.
\end{equation}

\smallskip In GTR, the electromagnetic field Eqs.(3.7), (3.8) and the
energy-momentum tensor $T_{n}^{m}$ of the electromagnetic field (3.10), have
the same form as below, the important difference being that in RTG all
fields depend on the coordinates in the underlying Minkowski space-time.

\section{ The electrogravitational field produced by a charged mass point in
RTG}

In the framework of GRT the problem of finding the field produced by a
charged mass point with mass $m$ and electric charge $q$, was solved by
Nordstr\"{o}m and by Jeffrey (see for example Wang 1979, Section 56).

The problem is static and spherically symmetric. The nonzero components of
the Riemannian metric which represent the electrogravitational field, are
the following:

\begin{equation}
g_{11}=-\frac{1}{1-\frac{2m}{r}+\frac{q^{2}}{r^{2}}}\mbox{\rm  , }%
g_{22}=-r^{2}\mbox{\rm  , }g_{33}=-r^{2}\sin ^{2}\theta \mbox{\rm  , }%
g_{44}=1-\frac{2m}{r}+\frac{q^{2}}{r^{2}}.
\end{equation}
\quad In the above relations $r$ and $\theta $ are two of the spherical
coordinates \{$r,\theta ,\varphi ,t$\} centered in the charged mass point,
in which are written the components of the metric. The domains of
definition for these coordinates are: $0\leq r_{g}<r<\infty $ , $0\leq
\theta \leq \pi $ , $0\leq \varphi \leq 2\pi $ , $-\infty <t<\infty $, $%
r_{g} $ representing the gravitational radius of the point source. According
to GRT, the value of this gravitational radius depends on the relation
between $q^{2}$ and $m^{2}$ in the following manner:

\begin{equation}
r_{g}=\left\{ 
\begin{array}{c}
m+\sqrt{m^{2}-q^{2}}\mbox{\rm  , for }q^{2}\leq m^{2} \\ 
\mbox{\rm  \qquad \qquad }0\mbox{\rm  \quad \quad , for }q^{2}>m^{2}
\end{array}
\right.
\end{equation}

We observe that the function $r_{g}$ depending on $q^{2}$ and $m^{2}$ has a
discontinuity in $q^{2}=m^{2}$.

The static and spherically symmetric electrogravitational field (4.1) has
been obtained solving the coupled system of Einstein's Eqs. and
Maxwell's Eqs. (3.7), (3.8). In the selected system of coordinates the
magnetic field $B^{\alpha }$ is zero, an only the radial component $%
E_{1}=E(r)$ of the electric field is non-vanishing, depending only on the
coordinate $r.$ From (3.3) we can conclude that the only nonzero components
of the electromagnetic tensor and of its dual are:

\begin{equation}
F^{23}(r)=-F^{32}(r)=\hat{F}_{14}(r)=-\hat{F}_{41}(r)=E(r).
\end{equation}

The expression of the unknown function $E(r)$ has been obtained solving
(3.7), (3.8), with $J^{n}\equiv $ 0. This $E(r)$ has been introduced in the
expression (3.10), thus getting the energy-momentum tensor of the considered
electromagnetic field. Finally, solving Einstein's Eqs. the solution (4.1)
has been obtained.

Let us now consider the problem of finding the electrogravitational field
produced by a charged mass point according to RTG. We must solve the system
of Eqs. (2.3), (2.4), (3.7), (3.8) in terms of the coordinate of the
underlying Minkowski space-time. Only those solutions that satisfy CP can
represent the physical acceptable fields.

I suppose that the spherical coordinates \{$r,\theta ,\varphi ,t$\} centered
in the charged mass point are the coordinates in the underlying Minkowski
space-time. The solution (4.1) satisfies Eqs. (2.3) (without the mass terms), (2.4), (3.7), (3.8). I
verify now if this solution satisfies (2.4) .

The metric of the Minkowski space-time in which we happen to be when the
gravitational field is switched off is:

\begin{equation}
d\sigma ^{2}=dt^{2}-dr^{2}-r^{2}d\theta ^{2}-r^{2}\sin ^{2}\theta d\varphi
^{2}.
\end{equation}

From (4.1), the nonzero components of the tensor $g^{mn}$ are:

\begin{equation}
g^{11}=-\left( 1-\frac{2m}{r}+\frac{q^{2}}{r^{2}}\right) \mbox{\rm  , }%
g^{22}=-\frac{1}{r^{2}}\mbox{\rm  , }g^{33}=-\frac{1}{r^{2}\sin ^{2}\theta }%
\mbox{\rm  , }g^{44}=\frac{1}{1-\frac{2m}{r}+\frac{q^{2}}{r^{2}}},
\end{equation}
and the determinant of the metric tensor $g_{mn}$ is given as follows:

\begin{equation}
g=-r^{4}\sin ^{2}\theta .
\end{equation}

From (4.5), (4.6) and (2.2) we obtain the nonzero components of $\tilde{g}%
^{mn}$:

\begin{equation}
\tilde{g}^{11}=-\sin \theta \left( r^{2}-2mr+q^{2}\right) \mbox{\rm  , }%
\tilde{g}^{22}=-\sin \theta \mbox{\rm  , }\tilde{g}^{33}=-\frac{1}{\sin
\theta }\mbox{\rm  , }\tilde{g}^{44}=\frac{r^{2}\sin \theta }{1-\frac{2m}{r}+%
\frac{q^{2}}{r^{2}}}.
\end{equation}

Taking into account (4.4), the coefficients of the metric connection
generated by \thinspace $\gamma _{mn}$ are:

\begin{equation}
\gamma _{22}^{1}=-r\mbox{\rm  , }\gamma _{33}^{1}=-r\sin ^{2}\theta %
\mbox{\rm  , }\gamma _{12}^{2}=\gamma _{13}^{3}=\frac{1}{r}\mbox{\rm  , }%
\gamma _{33}^{2}=-\sin \theta \cos \theta \mbox{\rm  , }\gamma
_{23}^{3}=\cot \theta .
\end{equation}

According to (4.7), (4.8), making $n=1$ in the system (2.4'), we obtain an
equation which is not fulfilled.

So, the solution (4.1) written in the spherical coordinates \{$r,\theta
,\varphi ,t$\} with the metric tensor of the Minkowski space-time having the
form (4.4), does not satisfy Eq. (2.4); hence it is not an admissible
solution in RTG.

For finding an admissible solution in RTG, I use the same procedure as
Logunov \& Mestvirishvili (1989), Chapter 13. Thus, I am looking for a
system of coordinates \{$\eta ^{i}$\}$=\{R,$ $\Theta ,$ $\Phi ,$ $T\}$ in
which Eqs. (2.3) (without mass terms), (3.7), (3.8) are fulfilled, Eqs. (2.4) establishing a one
to one relationship with a nonzero Jacobian, between the sets of coordinates
\{$\eta ^{i}$\} and \{$\xi ^{i}$\}=\{$r,$ $\theta ,$ $\varphi ,$ $t$\}, in
the Minkowski space-time. I shift from the variables \{$\xi ^{i}$\} to the
variables \{$\eta ^{i}$\} assuming that:

\begin{equation}
R=R(r),\mbox{\rm  }\Theta =\theta ,\mbox{\rm  }\Phi =\varphi ,\mbox{\rm  }%
T=t.
\end{equation}
The function $R(r)$ must satisfy the following restrictions:

\begin{equation}
R(r)>0\mbox{\rm  , for }r>r_{g}\mbox{\rm  \quad , \quad }\frac{dR}{dr}>0%
\mbox{\rm  , for  }r>r_{g}
\end{equation}
The transformation is made in such a way that when the gravitational field
disappears the metric of the underlying Minkowski space-time takes the form:

\begin{equation}
d\sigma ^{2}=dt^{2}-dR^{2}-R^{2}d\theta ^{2}-R^{2}\sin ^{2}\theta d\varphi
^{2}.
\end{equation}

The system of Eqs. (2.4) makes possible the finding of the explicit form of
the function $R(r)$. Eqs. (2.4) can be written in the following form (see
the relations (13.17), (13.22), from Logunov \& Mestvirishvili 1989):

\begin{equation}
\frac{1}{\sqrt{-g(\xi )}}\frac{\partial }{\partial \xi ^{m}}\left( \sqrt{%
-g(\xi )}g^{mn}(\xi )\frac{\partial \eta ^{p}}{\partial \xi ^{n}}\right)
=-\gamma _{mn}^{p}(\eta )\frac{\partial \eta ^{m}}{\partial \xi ^{i}}.\frac{%
\partial \eta ^{n}}{\partial \xi ^{j}}g^{ij}(\xi ).
\end{equation}

From (4.11) the nonzero components $\gamma _{mn}^{p}(\eta )$ have the
expressions:

\begin{equation}
\gamma _{22}^{1}=-R\mbox{\rm  , }\gamma _{33}^{1}=-R\sin ^{2}\theta %
\mbox{\rm  , }\gamma _{12}^{2}=\gamma _{13}^{3}=\frac{1}{R}\mbox{\rm  , }%
\gamma _{33}^{2}=-\sin \theta \cos \theta \mbox{\rm  , }\gamma
_{23}^{3}=\cot \theta .
\end{equation}

For $n=2,3,4$ taking into account (4.5), (4.6), (4.13), Eqs. of the system
(2.4') become identities, while for $n=1$ Eq. assumes the form:

\begin{equation}
\frac{d^{2}R(r)}{dr^{2}}(r^{2}-2mr+q^{2})+2(r-m)\frac{dR(r)}{dr}-2R(r)=0%
\mbox{\rm  , for }r>r_{g}.
\end{equation}
Here the gravitational radius $r_{g}$ is given by the expressions (4.2).

The solution of this Eq. depends essentially on the relation between $q^{2}$
and $m^{2}$ (see So\'{o}s and me 2000, Section 4).

\smallskip

\section{Case {\bf $q^{2}\leq m^{2}$}}

\smallskip

The solution of Eq. (4.14) obtained in this case is the following (see So\'{o}s and me
2000, Section 4):

\begin{equation}
R(r)=r-m\mbox{\rm  , for }r>r_{g}=m+\sqrt{m^{2}-q^{2}}
\end{equation}
the admissible minimum value for $R(r)$ being:

\begin{equation}
R_{g}=\sqrt{m^{2}-q^{2}}.
\end{equation}
$R_{g}$ represents the gravitational radius of our point source in the
system of coordinates \{$\eta ^{i}$\}.

From the tensor transformation law, the nonzero components of the effective
Riemannian metric $g_{mn}$ in the new coordinate system \{$\eta ^{i}$\}=\{$%
R,\theta ,\varphi ,t$\} are:

\begin{equation}
g_{11}=-\frac{(R+m)^{2}}{R^{2}-m^{2}+q^{2}}\mbox{\rm  , }g_{22}=-(R+m)^{2}%
\mbox{\rm  , }g_{33}=-(R+m)^{2}\sin ^{2}\theta \mbox{\rm  , }g_{44}=\frac{%
R^{2}-m^{2}+q^{2}}{(R+m)^{2}}.
\end{equation}

The above solution was obtained for the first time by Karabut \& Chugreev
(1987); E.So\'{o}s and me (2000) have obtained the same result.

The analysis of the problem in the framework of RTG doesn't stop here. A
necessary condition for that the solution (5.3) has physical sense, is CP.
Allowing for the expression (4.11) of the underling Minkowski space-time, I
choose the Minkowski isotropic vector $u=(0,1,0,R(r)).$ The condition (2.6)
becomes:

\begin{equation}
(R^{2}-m^{2}+q^{2})R^{2}\leq (R+m)^{4},\mbox{\rm  \quad for \quad }R>R_{g},
\end{equation}
$R_{g}$ being given by (5.2). It is easy to verify that this restriction is
fulfilled.

Also, for the Minkowski isotropic vector $u=(1,0,0,1)$, the causality
condition:

\begin{equation}
(R^{2}-m^{2}+q^{2})^{2}\leq (R+m)^{4},\mbox{\rm  \quad for \quad }R>R_{g},
\end{equation}
must be satisfied. Because (5.4) is valid, the inequality (5.5) is obviously
satisfied.

So, the solution (5.3) represents the physically acceptable solution in RTG.

\smallskip

\section{Case {\bf $q^{2}>m^{2}$}}

\smallskip

According to the empirical values of $q$ and $m,$ for a single electron just
this case is true. So, it's important to see how looks the solution in this
case. This case was firstly analysed in RTG, by So\'{o}s and by me in (2000).

Integrating Eq. (4.14) and choosing one of the two constants of integration
which appear, in such way that for $r$ tends to $\infty $ $,$ $\frac{R(r)}{r}
$ tends to 1, we get:

\begin{equation}
R(r)=r-m+C\left( 1-\frac{r-m}{p}\arctan \frac{p}{r-m}\right) 
\mbox{\rm  ,
for }r>r_{g},
\end{equation}
where $p=\sqrt{q^{2}-m^{2}}$ and $C$ is a real constant.

The components of the effective Riemannian space-time , in the system of
coordinates $\{\eta ^{i}\}$ are the following:

\begin{equation}
g_{11}=\left( \frac{dr(R)}{dR}\right) ^{2}\left( -\frac{1}{1-\frac{2m}{r(R)}+%
\frac{q^{2}}{r^{2}(R)}}\right) \mbox{\rm  ,  }g_{22}=-r^{2}(R)\mbox{\rm  ,  }%
g_{33}=-r^{2}(R)\sin ^{2}\theta \mbox{\rm   ,  }g_{44}=1-\frac{2m}{r(R)}+%
\frac{q^{2}}{r^{2}(R)}
\end{equation}
the function $r(R)$ being implicitly given by (6.1).

Eqs. (2.4) being general covariant, the system of coordinates $\{\eta ^{i}\}$
is not a privileged one. Thus, the solution (4.1) is also a solution of this
system of Eqs., but the system of Minkowski coordinates in which is written
this solution, is one in which the Minkowskian metric does not have the form
(4.4) but the form:

\begin{equation}
d\sigma ^{2}=dt^{2}-\left( \frac{dR(r)}{dr}\right)
^{2}dr^{2}-R^{2}(r)d\theta ^{2}-R^{2}(r)\sin ^{2}\theta d\varphi ^{2},
\end{equation}
$R(r)$ being explicitly given by (6.1).

Therefore, the solution obtained in this case can be written either in the
form (6.2) in the system of coordinates for which the underlying Minkowskian
metric has the form (4.11), or in the form (4.1) in the system of
coordinates for which the underlying Minkowskian metric is (6.3).  
So\'{o}s E. and me (2000) have considered the form (6.2), (4.11).

Apparently, our problem has a family of solutions depending on the parameter 
$C$. Any way, the obtained solutions must satisfy CP in RTG. Because we
don't have the explicit form of the inverse function $r(R)$, I verify if
this principle is satisfied, using the form (4.1), (6.3) of the solutions.

Considering the Minkowskian isotropic vector $u=(0,1,0,R(r)),$the causality
condition (2.6) becomes:

\begin{equation}
\left( r^{2}-2mr+q^{2}\right) R^{2}(r)\leq r^{4}\mbox{\rm  \quad for }r>r_{g}%
\mbox{\rm  , }R>R_{g}\mbox{\rm  , }
\end{equation}
where $R_{g}\geq 0$ is the lower admissible bound for the function $R(r)$.
This value must be also determined.

Also, for the Minkowskian isotropic vector $u=\left( 1,0,0,\frac{dR(r)}{dr}%
\right) $ the following inequality must be valid:

\begin{equation}
\left( r^{2}-2mr+q^{2}\right) ^{2}\left( \frac{dR(r)}{dr}\right) ^{2}\leq
r^{4}\mbox{\rm  \quad for }r>r_{g}\mbox{\rm  , }R>R_{g}\mbox{\rm  . }
\end{equation}

For the considered case $q^{2}>m^{2}$, according to (4.2), $r_{g}=0.$ I'll
show that if we want the causality conditions (6.4), (6.5) be satisfied, $%
r_{g}$ can't be zero. So, the solution obtained by  So\'{o}s and by me (2000)
doesn't fulfill CP in RTG. This is the main reason for what I have
reanalysed the problem considered by us.

Indeed, if $r_{g}=0$ and in (6.4) we make $r$ tends to zero and we get $%
R(0)=0.$ This yields:

\begin{equation}
C=\frac{m}{1-\frac{m}{p}\arctan \frac{p}{m}}\equiv C_{1.}
\end{equation}
And if in (6.5) we make $r$ tends to zero, we find $\frac{dR(0)}{dr}=0,$
implying:  

\begin{equation}
C=\frac{1}{-\frac{1}{p}\arctan \frac{p}{m}+\frac{m}{m^{2}+p^{2}}}\equiv
C_{2.}
\end{equation}

It's easy to see that $C_{1}<C_{2}.$ Hence, we obtain a contradiction, since
from (6.6), (6.7) we must have $C_{1}=C_{2}$. It results that the restrictions (6.4),
(6.5) could be fulfilled only if :

\begin{equation}
r_{g}>0.
\end{equation}

Now, I am looking for the value of this $r_{g}.$

I return to the conditions which must be fulfilled by the function $R(r).$
The function $R(r)$ has the analytical expression (6.1), and must satisfy
the conditions (4.10), (6.4), (6.5). The real constant $C$ will be
determined such that the positive and increasing function $R(r)$ gets in the
possible minimum value $r=r_{g}$ its possible minimum value denoted by $%
R_{g}.$

From the analytical expression (6.1) of $R(r),$ we notice that this function
has the straight line $R(r)=r-m$ like oblique asymptote at $\infty .$
Allowing for the fact that the following function is positive:

\begin{equation}
f(r)=1-\frac{r-m}{p}\arctan \frac{p}{r-m}\geq 0\mbox{\rm  , for }r>0,
\end{equation}
from (6.1) we get :

\begin{equation}
R(r)\leq r-m\mbox{\rm  \qquad if and only if \qquad }C\leq 0
\end{equation}
and

\begin{equation}
R(r)>r-m\mbox{\rm  \qquad if and only if \qquad }C>0.
\end{equation}

I'll present separately the two possibilities (6.10), (6.11), respectively.

a) $C\leq 0$

In this case, taking into account the expression (6.1) of the function $%
R(r), $ the condition (4.10)$_{1}$ which this function must satisfy and the
relation (6.9), we conclude that:

\begin{equation}
r>m.
\end{equation}

Deriving the function (6.1), we get:

\begin{equation}
\frac{dR(r)}{dr}=1+C\left( -\frac{1}{p}\arctan \frac{p}{r-m}+\frac{r-m}{%
(r-m)^{2}+p^{2}}\right) .
\end{equation}

Since, the following function is negative:

\begin{equation}
h(r)=-\frac{1}{p}\arctan \frac{p}{r-m}+\frac{r-m}{(r-m)^{2}+p^{2}}\leq 0%
\mbox{\rm  , for }r>0,
\end{equation}
in the considered case we obtain:

\begin{equation}
\frac{dR(r)}{dr}\geq 1\mbox{\rm  , for }r>r_{g}.
\end{equation}

So, the condition (4.10)$_{2}$ is obviously fulfilled.

The causality condition (6.5) must be also valid for $r>r_{g},$ $R>R_{g}$,
and allowing for (6.15) we find:

\begin{equation}
r\geq \frac{q^{2}}{2m}.
\end{equation}

In view of (6.10), (6.16) and of the fact that $r-m<r$, the causality
condition (6.4) is satisfied.

The conditions (6.12), (6.16)are necessary conditions. For that (4.10)$_{2}$
and (6.5) to be valid, the real constant $C$ would have to satisfy:

\begin{equation}
C>\frac{m-r}{1-\frac{r-m}{p}\arctan \frac{p}{r-m}}\mbox{\rm  , for }r>r_{g},
\end{equation}
and respectively:

\begin{equation}
C\geq \frac{2mr-q^{2}}{\left( r^{2}-2mr+q^{2}\right) \left( -\frac{1}{p}%
\arctan \frac{p}{r-m}+\frac{r-m}{(r-m)^{2}+p^{2}}\right) }\mbox{\rm  , for }%
r>r_{g}.
\end{equation}

According to (6.12), (6.16), we find:

\begin{equation}
r_{g}=\left\{ 
\begin{array}{c}
\mbox{\rm  \quad }m\mbox{\rm  ,\quad for }m^{2}<q^{2}<2m^{2} \\ 
\frac{q^{2}}{2m}\mbox{\rm  , \quad for }2m^{2}\leq q^{2}\mbox{\rm  .}
\end{array}
\right.
\end{equation}

Taking into account the criterion for choosing the real constant $C$, the
inequalities (6.17), (6.18), where $r_{g}$ has the value (6.19), we finally
get:

\begin{equation}
C=0.
\end{equation}

Substituting (6.20) into (6.1), we find in this case :

\begin{equation}
R(r)=r-m\mbox{\rm  , for }r>r_{g}.
\end{equation}
$r_{g}$ being given by (6.19).

b) $C>0$

In this case, $r$ also can not take values smaller than $m$. Indeed, if $r$
would take the values smaller than $m$, from (6.13), the point $r=m$ would
be a return point for $R(r),$ so, $R(r)$ wouldn't be strictly increasing
function.

Besides, in this case the condition (4.10)$_{2}$ is not fulfilled. Indeed,
from (6.13), allowing for the fact that the function $h(r)$ from (6.14) is
monotonously increasing function, and tends to zero for $r$ tends to $\infty
,$ we get that there exist $r>r_{g}\geq m$ such that $\frac{dR(r)}{dr}=0.$

So, for $C>0$, the function $R=R(r)$ can't be strictly increasing function
for $r>r_{g}.$

Summing up, if the relation between $q^{2}$and $m^{2}$ is $q^{2}>m^{2}$, the
only function, in the form (6.1), which satisfies the restrictions (4.10),
(6.4), (6.5) is the function (6.21), where $r_{g}$ is given by (6.19).

Then the analytical expression of $R=R(r)$ is the same as in the case $%
q^{2}\leq m^{2}$, but its domain of definition is different.

\smallskip

\section{Conclusions}

\smallskip

We can conclude that the solution of our problem according to RTG is unique
for all order relations which can exist between $m^{2}$ and $q^{2}$. The
analytical expression of the solution doesn't depend on the relation between 
$m^{2}$ and $q^{2}$, but its domain of definition, i.e. $r_{g}$, depends on
this relation. On the basis of (5.1), (6.21), (6.19), we can write:

\begin{equation}
R(r)=r-m\mbox{\rm  , for }r>r_{g}.
\end{equation}
where

\begin{equation}
r_{g}=\left\{ 
\begin{array}{c}
m+\sqrt{m^{2}-q^{2}}\mbox{\rm  , for }q^{2}\leq m^{2} \\ 
m\mbox{\rm  , for }m^{2}<q^{2}<2m^{2} \\ 
\frac{q^{2}}{2m}\mbox{\rm  , for }2m^{2}\leq q^{2}.
\end{array}
\right.
\end{equation}

We notice that the function $r_{g}$ depending on $q^{2}$ and $m^{2}$ is a
continuous one.

The expression of the effective Riemannian metric has in the system of
coordinates $\{\xi ^{i}\}=\{r,\theta ,\varphi ,t\}$ the same form (4.1) like
in GRT. But in RTG , these coordinates are the spatial-temporal coordinates
in the Minkowski universe. The line element of this universe doesn't have
the form (4.4) but the form (6.3). Substituting (7.1) into (6.3), we get
this form:

\begin{equation}
d\sigma ^{2}=dt^{2}-dr^{2}-(r-m)^{2}d\theta ^{2}-(r-m)^{2}\sin ^{2}\theta
d\varphi ^{2}.
\end{equation}

Comparing (4.2) and (7.2), I also stress the difference between the
gravitational radius in this two theories.

We can also write the components of the effective Riemannian metric in the
system of coordinates $\{\eta ^{i}\}=\{R,\theta ,\varphi ,t\},$in which the
Minkowskian line element has the form (4.11). These are:

\begin{equation}
g_{11}=-\frac{(R+m)^{2}}{R^{2}-m^{2}+q^{2}}\mbox{\rm  , }g_{22}=-(R+m)^{2}%
\mbox{\rm  , }g_{33}=-(R+m)^{2}\sin ^{2}\theta \mbox{\rm  , }g_{44}=\frac{%
R^{2}-m^{2}+q^{2}}{(R+m)^{2}}.
\end{equation}

From (7.1), (7.2), the gravitational radius in the system of coordinates $%
\{\eta ^{i}\}$ has the expression:

\begin{equation}
R_{g}=\left\{ 
\begin{array}{c}
\sqrt{m^{2}-q^{2}}\mbox{\rm  , for }q^{2}\leq m^{2} \\ 
0\mbox{\rm  , for }m^{2}<q^{2}<2m^{2} \\ 
\frac{q^{2}-2m^{2}}{2m}\mbox{\rm  , for }2m^{2}\leq q^{2}.
\end{array}
\right.
\end{equation}
$R_{g}$ like function depending on $q^{2}$ and $m^{2},$ is also a continuous
function.

The obtained result shows again the important role played by CP in RTG:
indeed the analytical expression of the function $R=R(r)$ is much more
simple as that obtained by E So\'{o}s and by me (2000).

\smallskip

{\bf Acknowledgements. }I would like to thank my teacher E. So\'{o}s, who
has directed and guided me in this field, for many helpful discussions and
critical remarks which made me to undertake and finish this work.
I am also grateful to Prof. V. Soloviev for translating this paper into Russian.

\smallskip

\smallskip

\smallskip


\smallskip

\smallskip

\smallskip

\begin{thebibliography}{9}
\bibitem{1}  Ionescu, D. \& So\'{o}s, E. 2000 Electrogravitational Field
Produced by a Charged Mass Point in RTG. {\it Rev. Roum. Math. P. Appl.},
{\bf 45} (2), 251-260, 2000.

\bibitem{2}  Karabut, P. V. \& Chugreev Iu. V. 1987 Exterior Axial-Symmetric
Solution for a Spinning Charged Body in RTG. {\it Institute of High-Energy Physics
Preprint}, 87-142. Serpuhov.

\bibitem{3}  Logunov, A. A. \& Mestvirishvili, M. 1989 {\it \ The Relativistic Theory
of Gravitation}. Moscow: Mir.

\bibitem{4}  Logunov, A. A. 1997 {\it Relativistic Theory of Gravity and Mach
Principle}. Dubna.

\bibitem{5}  Moller, C. 1972 {\it The Theory of Relativity}. Oxford: Clarendon Press.

\bibitem{6}  So\'{o}s, E. 1992 {\it G\'{e}om\'{e}trie et \'{e}lectromagnetisme}.
University of Timisoara, Math. Dept..

\bibitem{7}  Truesdell, C. \& Toupin, R. A. 1960 {\it The Classical Field Theories}.
Handbuch der Physik, {\bf III/1}, pp. 226-858. Berlin, Gottingen, Heidelberg: Springer. 

\bibitem{8}  Wang, C. C. 1979 {\it Mathematical Principles of Mechanics and
Electromagnetism, Part B: Electromagnetism and Gravitation}. New York, London: Plenum Press.




\end{thebibliography}
\end{document}